\documentclass[aps,prb,twocolumn,showpacs,superscriptaddress,floatfix]{revtex4-1}  
\usepackage{caption}
\usepackage{graphicx} 
\usepackage{subfigure}  
\usepackage{dcolumn}
\usepackage{bm}
\usepackage{epstopdf} 
\usepackage{amsmath}      
\usepackage{amssymb}
\begin{document}

\title{An \textit{Ab Initio} Description of the Mott Metal-Insulator Transition of M$_{2}$ Vanadium Dioxide}
\author{J.~M.~Booth}
\email{jamie.booth@rmit.edu.au}
\affiliation{Australian Research Council Centre of Excellence for Exciton Science, School Science, RMIT University, Melbourne 3001, VIC, Australia}
\affiliation{Theoretical Chemical and Quantum Physics, School of Science, RMIT University, Melbourne VIC 3001, Australia}

\author{D.~W.~Drumm}
\affiliation{Australian Research Council Centre of Excellence for Nanoscale BioPhotonics, School of Science, RMIT University, Melbourne 3001, VIC, Australia}

\author{P.~S.~Casey}
\affiliation{CSIRO Materials, Clayton VIC 3168, Australia}

\author{S.~K.~Bhargava}
\affiliation{Centre for Advanced Materials and Industrial Chemistry, School Science, RMIT University, Melbourne VIC 3001, Australia}

\author{J.~S.~Smith}
\affiliation{Theoretical Chemical and Quantum Physics, School of Science, RMIT University, Melbourne VIC 3001, Australia}

\author{S.~P.~Russo}
\affiliation{Australian Research Council Centre of Excellence for Exciton Science, School Science, RMIT University, Melbourne 3001, VIC, Australia}
\affiliation{Theoretical Chemical and Quantum Physics, School of Science, RMIT University, Melbourne VIC 3001, Australia}

\date{\today}

\begin{abstract}
Using an \textit{ab initio} approach based on the GW approximation which includes strong local \textbf{k}-space correlations, the Metal-Insulator Transition of M$_2$ vanadium dioxide is broken down into its component parts and investigated. Similarly to the M$_{1}$ structure, the Peierls pairing of the M$_{2}$ structure results in bonding-antibonding splitting which stabilizes states in which the majority of the charge density resides on the Peierls chain. This is insufficient to drop all of the bonding states into the lower Hubbard band however. An antiferroelectric distortion on the neighboring vanadium chain is required to reduce the repulsion felt by the Peierls bonding states by increasing the distances between the vanadium and apical oxygen atoms, lowering the potential overlap thus reducing the charge density accumulation and thereby the electronic repulsion. The antibonding states are simultaneously pushed into the upper Hubbard band. The data indicate that sufficiently modified GW calculations are able to describe the interplay of the atomic and electronic structures occurring in Mott metal-insulator transitions.
\end{abstract}

\maketitle

\section{Introduction}
M$_{1}$ Vanadium dioxide undergoes a transition from an insulating P2$_1$/c (14) monoclinic structure to a metallic tetragonal P4$_2$/mnm (136) structure at approximately 340 K.\cite{Morin1959,Goodenough1971} The 3d$^1$ electronic configuration results in strong electronic correlations in the metallic structure \cite{Tomczak2008} that drive the adoption of the insulating state as the structure cools. This transition has many useful properties such as its ultrafast timescale,\cite{Cavalleri2001} and the modulation of the critical temperature T$_c$ by doping \cite{Lawton1995,Imada1998,Wei2012} or inputting stress or strain \cite{Wei2009,Cao2009} which endow it with enormous promise for applications ranging from new transistor gates,\cite{Nakano2012} to ultrafast optical devices,\cite{Wall2012} and sensors. \cite{Zhou2008, Strelcov2009}

\begin{figure}[h!]
    \centering
    \subfigure[]{\includegraphics[width=0.9\columnwidth]{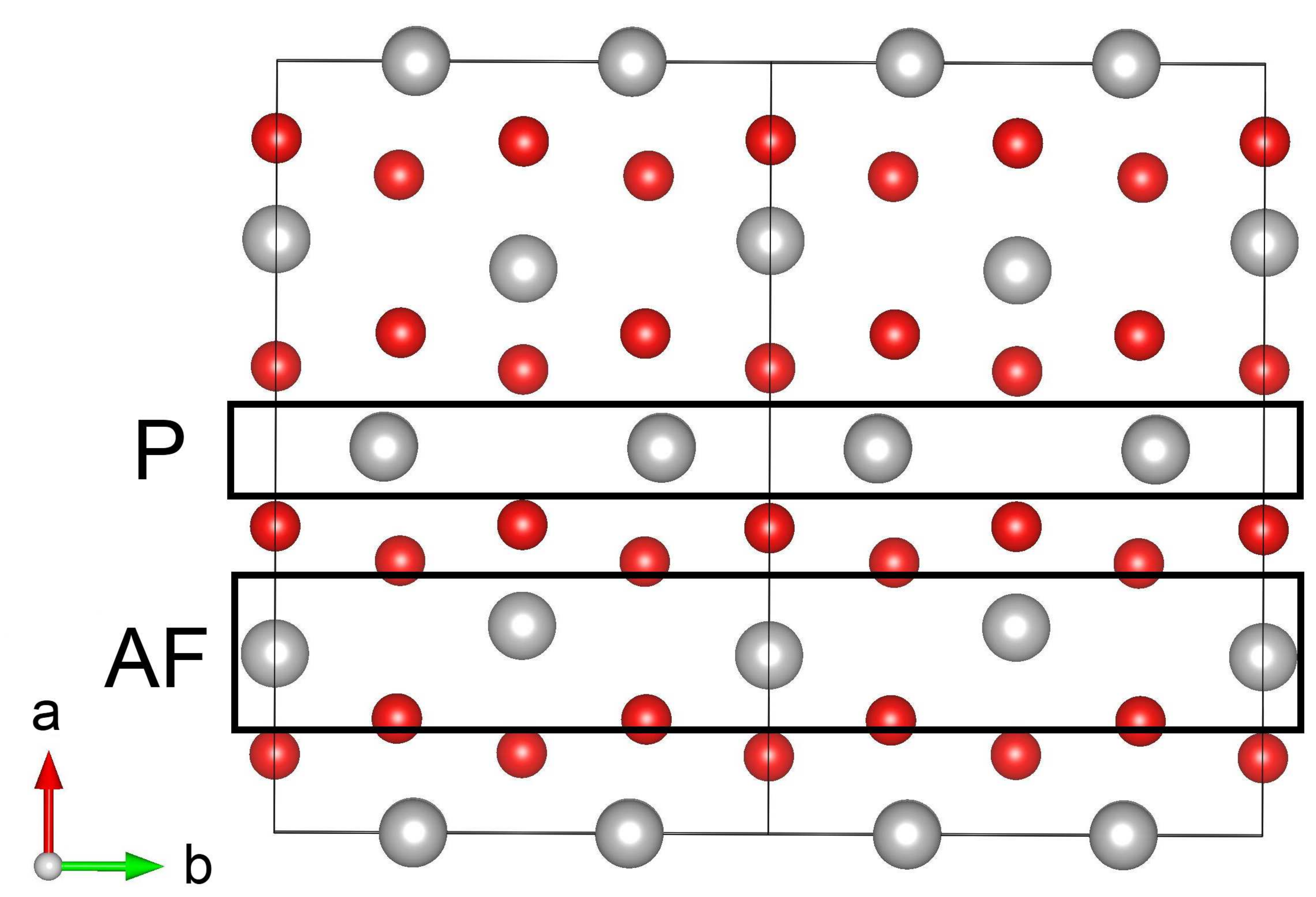}}\\
    \subfigure[]{\includegraphics[width=0.4\columnwidth]{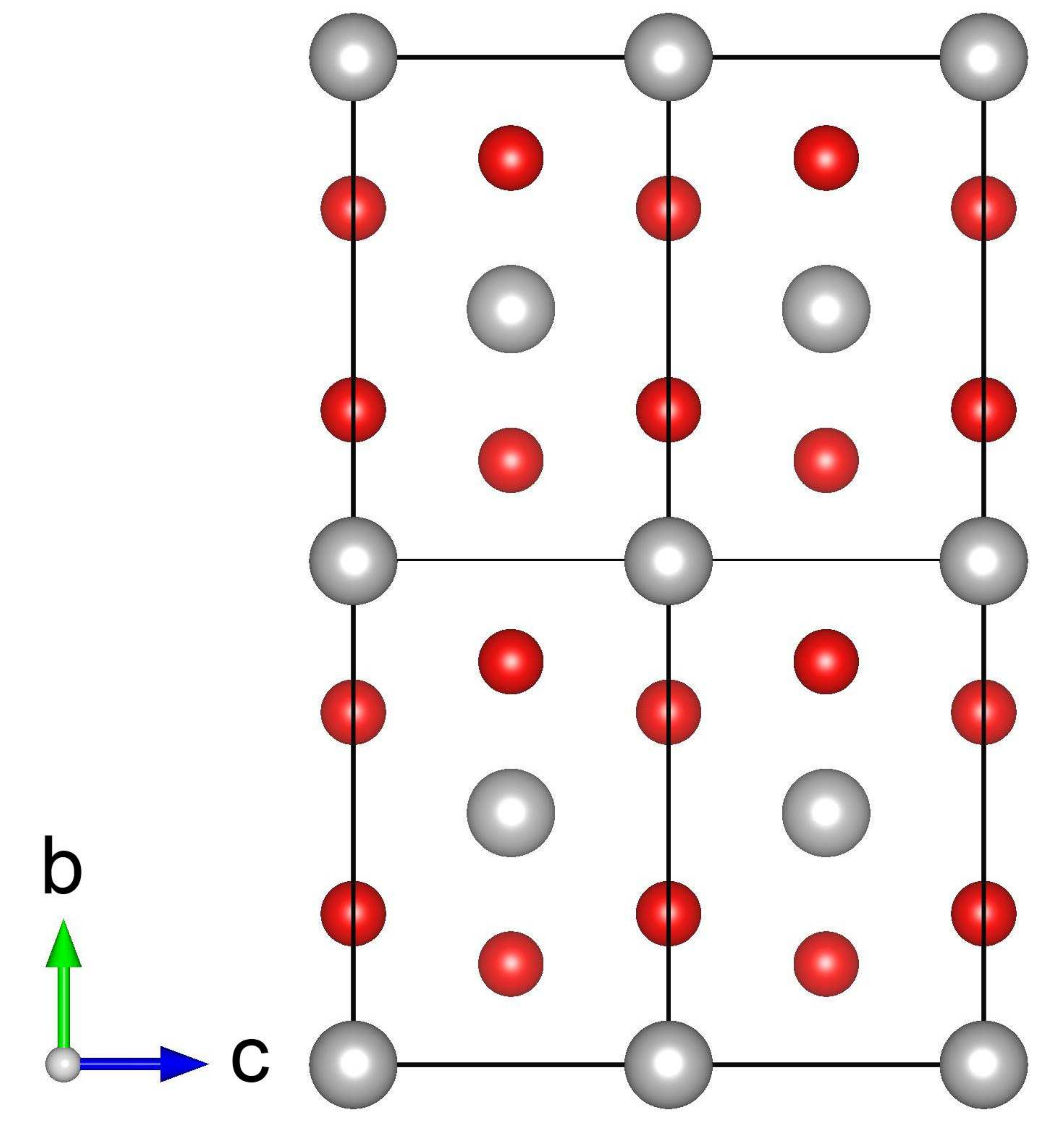}}\quad
    \subfigure[]{\includegraphics[width=0.55\columnwidth]{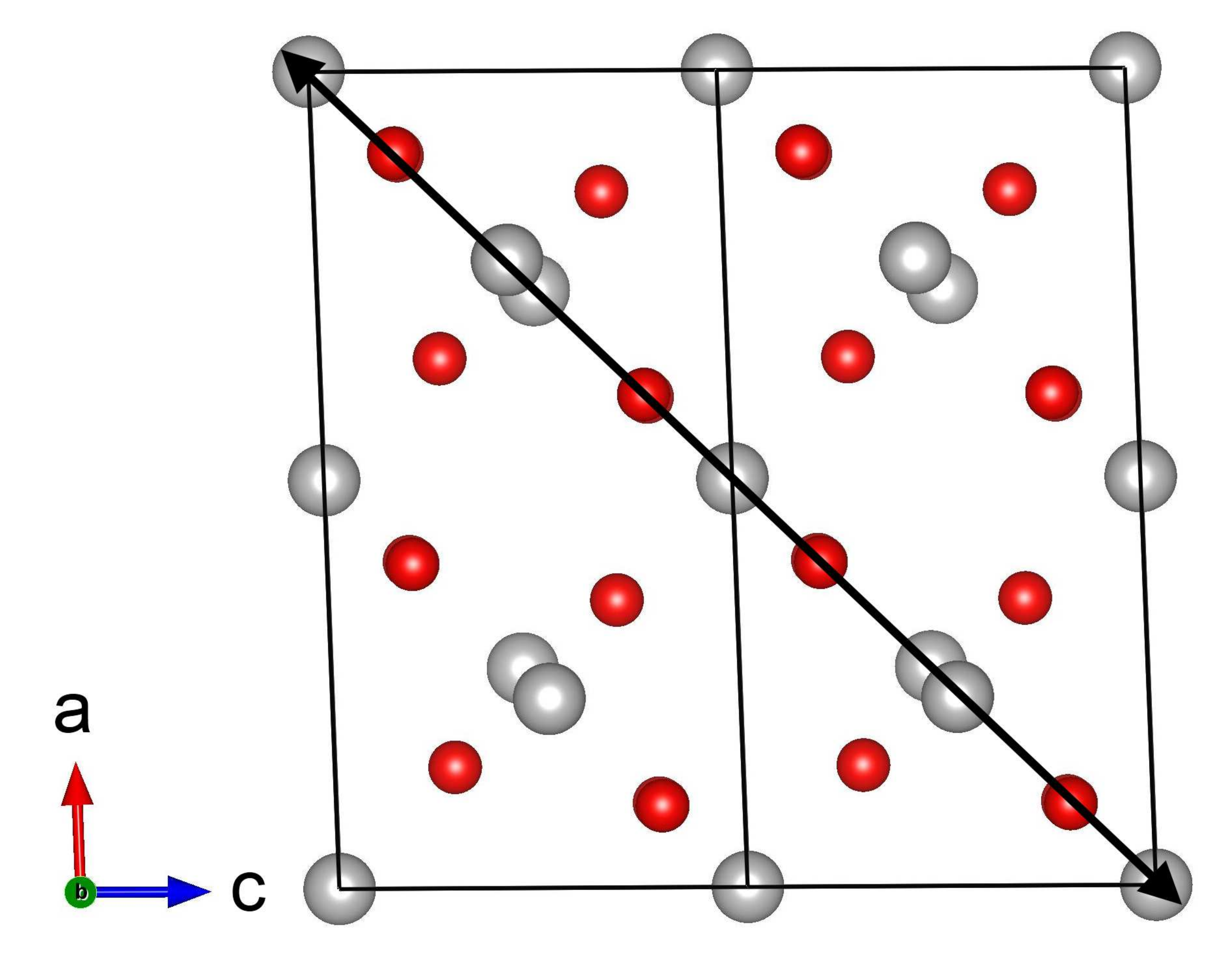}}
    \caption{\raggedright {a) View down $\langle001\rangle$ of the low temperature $C2/m$ monoclinic M$_2$ structure, vanadium atoms are gray and oxygen atoms are red. The alternating Peierls and Antiferroelectrically distorted chains are highlighted and labelled ``P" and ``AF" respectively, b) View down $\langle100\rangle$ of the high temperature tetragonal structure illustrating the evenly spaced vanadium chains and c) View down $\langle010\rangle$ of the monoclinic structure, the (201) plane is illustrated by a double-headed arrow and indicates the plane in which all charge densities are displayed.}}
\label{fig:schematics}
\end{figure}

The development of devices based on nanostructures of VO$_2$ however is complicated by the fact that the phase diagram of VO$_2$ is non-trivial. In doped systems (such as Cr-doped VO$_{2}$\cite{Pouget1976}) or systems under unaxial strain an M$_{2}$ \cite{Marezio1971} insulating structure forms. Its morphology differs from the M$_{1}$ in that rather than both vanadium chains which run along the tetragonal c-axis exhibiting Peierls pairing and an antiferroelectric twist, in the M$_{2}$ structure these chains alternate between Peierls paired but collinear, and antiferroelectrically distorted but evenly spaced (see Figure \ref{fig:schematics}). Studies on VO$_2$ nanobeams and nanowires in particular \cite{Zhang2009,Sohn2009,Jones2010,Guo2011} reveal that  stress and strain in nanobeam configurations commonly result in the appearance of the M$_{2}$ structure in the metal-insulator transition. Recent work by Park \textit{et al.} \cite{Park2013} identified a triple point between the tetragonal, M$_{1}$ and M$_{2}$ structures.

\begin{figure}[h!]
    \centering
    \includegraphics[width=\columnwidth]{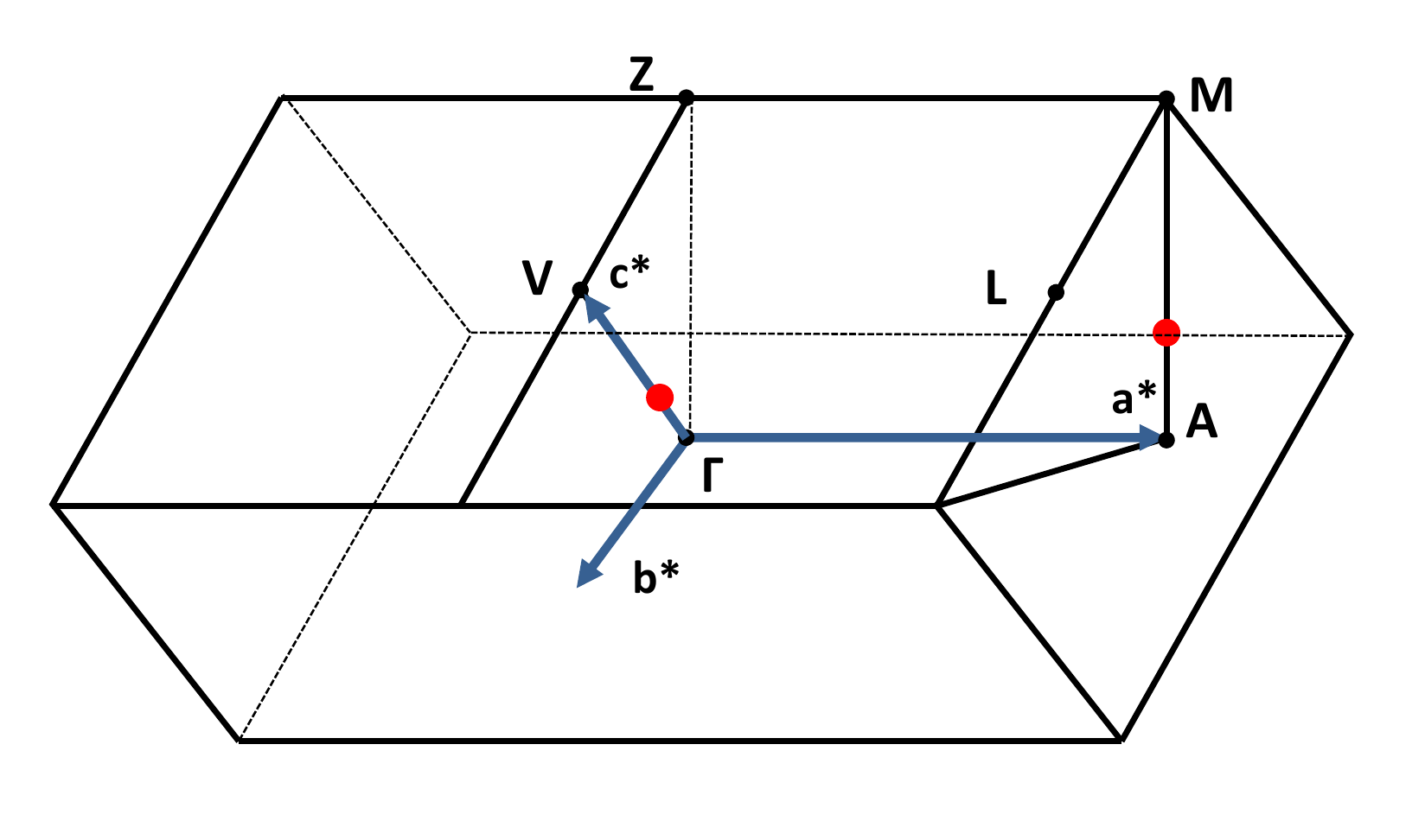}
    \caption{\raggedright {Brillouin zone of the C2/m monoclinic structure\cite{Bradley1972} of M$_{2}$ VO$_{2}$ used for all calculations in this work. The red circles mark the states used to construct the charge densities of Figure \ref{fig:crossover}.}}
\label{fig:BZ}
\end{figure}

However, while a wealth of information, both theoretical and experimental, exists concerning the M$_{1}$ to tetragonal transition, the M$_{2}$ structure's properties and dynamics have received considerably less attention. The recent focus on applications involving nanobeams and the study of Park \textit{et al.} \cite{Park2013} highlight the fact that this knowledge gap needs to be bridged. 

The most significant reason for this gap is that the M$_{2}$ form of VO$_2$ is a Mott Insulator,\cite{Pouget1974} and this renders many of the standard theoretical approaches to determination of the system's properties inapplicable. Density Functional Theory \cite{Kohn1965} in particular, while extremely successful when applied to weakly correlated systems, fails spectacularly when applied to Mott insulators due to its inability to correctly address non-local electron correlations. For example it commonly predicts Mott insulators such as CuO, CoO to be metals,\cite{Burke2012,Booth2016a} which renders it completely inappropriate for investigations on harnessing Mott insulating behavior. Despite this, Eyert \cite{Eyert2002} explored some of the properties of the electronic structure of M$_{2}$ VO$_2$ using DFT, and concluded that within the limits of the Local Density Approximation, the Peierls chain displayed character similar to the M$_{1}$ structure, while the antiferroelectric chain was rutile-like in electronic character. 

However the aforementioned lack of non-local correlations renders DFT a poor approximation, in particular the calculations were unable to open a band gap at the Fermi level in either the M$_{1}$ or M$_{2}$ phase, and thus the extact mechanism for the gap opening in M$_{2}$ VO$_2$ could not be rigorously determined. Other approaches have been developed to address non-local and strong electron correlations, such as the DFT+U \cite{Anisimov1997}, hybrid DFT functionals (such as HSE03 and B3LYP) which mix in an empirical amount of exact exchange,\cite{Heyd2003} and DFT+Dynamical Mean Field Theory,\cite{Kotliar2006} however their application to M$_{2}$ VO$_2$ is limited to the Hybrid Functional (HSE06) study of Eyert \cite{Eyert2011} and the DMFT studies of Tomczak \textit{et al.} \cite{Tomczak2008} and Brito \textit{et al.}\cite{Brito2016} These studies focused almost exclusively on the M$_{1}$ structure, and while the HSE data of Eyert revealed a gap in the band structure of M$_{2}$ VO$_2$, its origins are unclear in the context of the MIT and the empirical amount of exact exchange added by the functional, and in addition the splitting between the majority oxygen and majority metal $d$ orbital states was very different from that of the M$_{1}$ structure, suggesting issues with band-ordering in the hybrid approach.

The study of Tomczak \textit{et al.} concluded that the Peierls pairing of the M$_{2}$ structure is driven by strong correlations in the metallic state, analogously to M$_{1}$ VO$_2$, but did not address the electronic structure of the AF chain, and therefore did not illustrate the exact reason for the opening of a gap. The study of Brito \textit{et al.} confirmed the Mott insulating nature of the M$_{2}$ structure, and found a pole in the self-energy of paramagnetic M$_{2}$ in the $a_{1}g$ band (i.e. the states with density directed along the inter-vanadium axis) of the unpaired chain indicating a canonical Mott instability.

However, due to the aforementioned technological constraints, it has not been possible to clarify the significance of the displacive phase transition in the Mott transition of M$_{2}$ VO$_{2}$. That is, how do the atomic motions affect the stabilities of the electronic states, and how may they be influenced, particularly in nanobeam configurations in which stress and strain can be input. In a recent study\cite{Booth2016a} we demonstrated that strong local \textbf{k}-space correlations could be included in GW calculations by truncating the series expansion of the exponential operator in the evaluation of the exchange charge density in the limit of low \textbf{q}. Contrasting this method with standard G$_{0}$W$_{0}$ calculations revealed that the incomplete Peierls pairing of the M$_{2}$ structure when compared to M$_{1}$ results in band theory predicting a metallic structure. Specifically the antiferroelectrically distorted chain, which does not experience Peierls pairing results in standard GW predicting a ``quasiparticle" peak comprised of the 3d$_{z^{2}-r^{2}}$ states sitting at the Fermi level. 

Including strong correlations in the calculations split these states into the upper and lower Hubbard bands, which combined with the bonding and antibonding bands created by the Peierls distortion. However, while this calculation confirmed that the M$_{2}$ structure is indeed a Mott insulator, it did not explore the interplay between the structural rearrangements occurring across the transition, and the electronic structure. For the purposes of materials scientists creating devices based upon the MIT of VO$_{2}$, particularly those based on nanobeams in which the M$_{2}$ structure is commonly observed, this information is vital. In this work we address this by breaking the M$_{2}$ structural transition down into component parts and use the aforementioned adaptation of the GW approximation\cite{Booth2016a} to examine the electronic structure.

\begin{figure*}[h!]
    \centering
    \subfigure[]{\includegraphics[width=1.3\columnwidth]{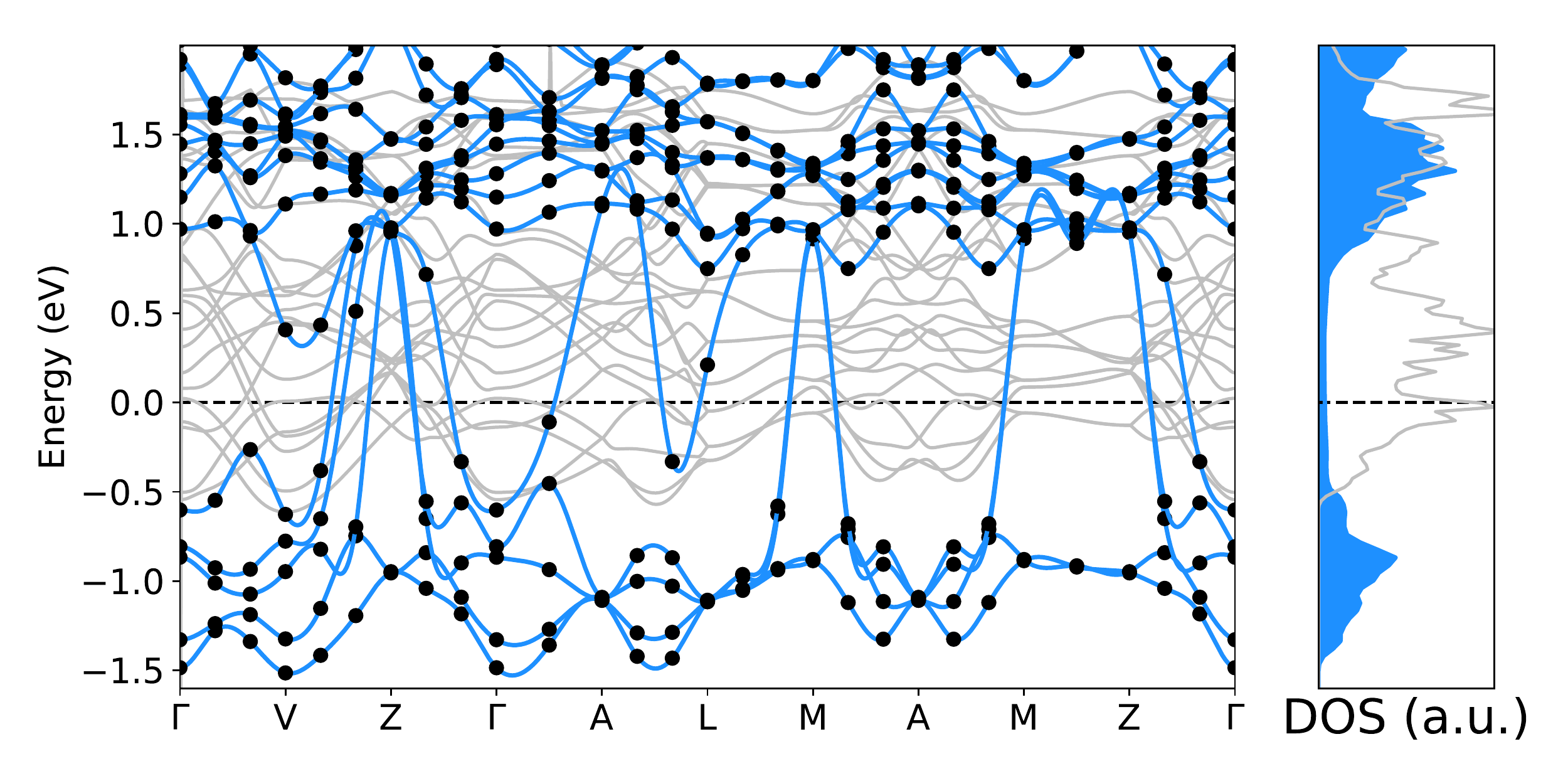}}
    \subfigure[]{\includegraphics[width=1.3\columnwidth]{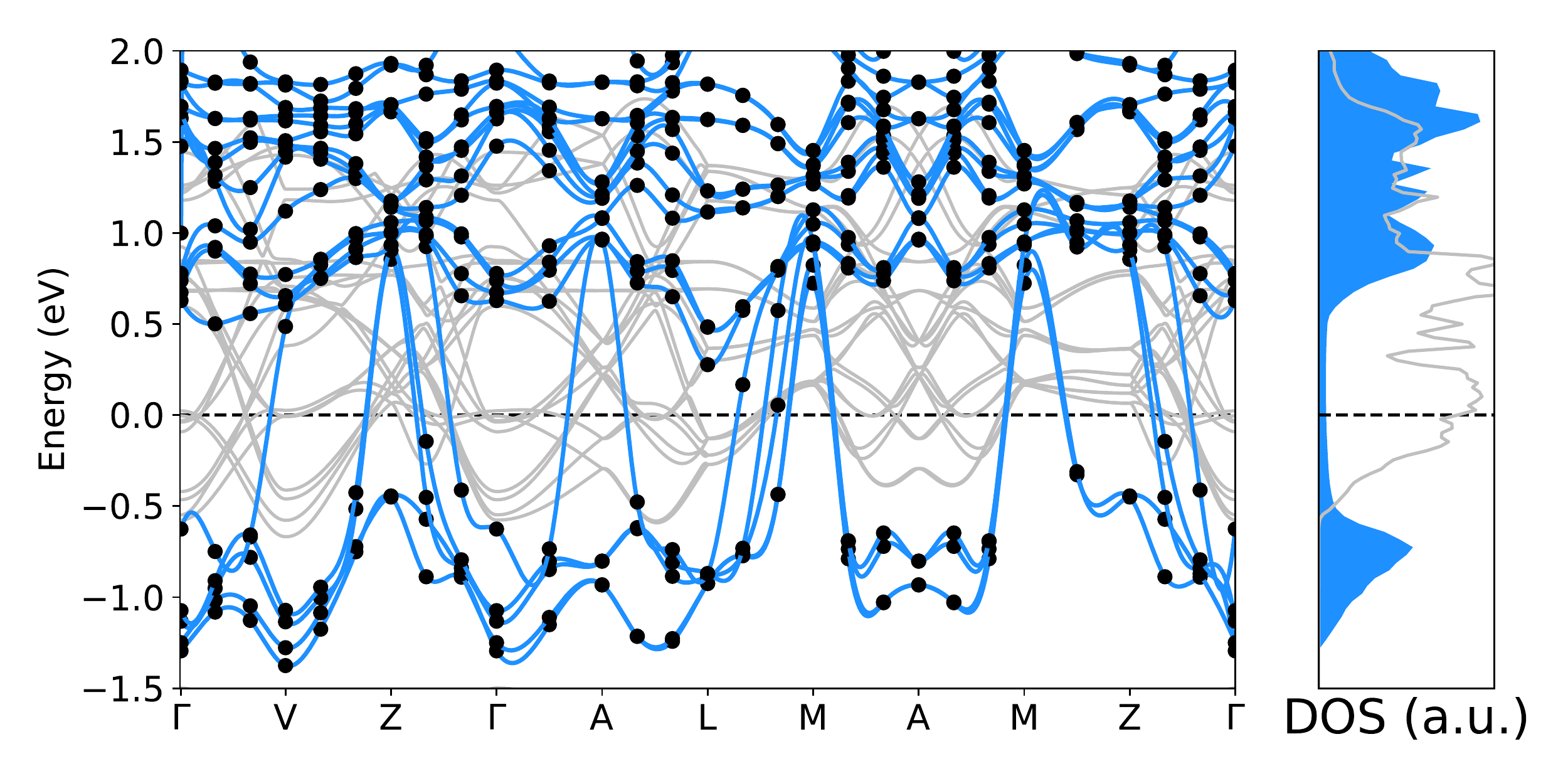}}
    \subfigure[]{\includegraphics[width=1.3\columnwidth]{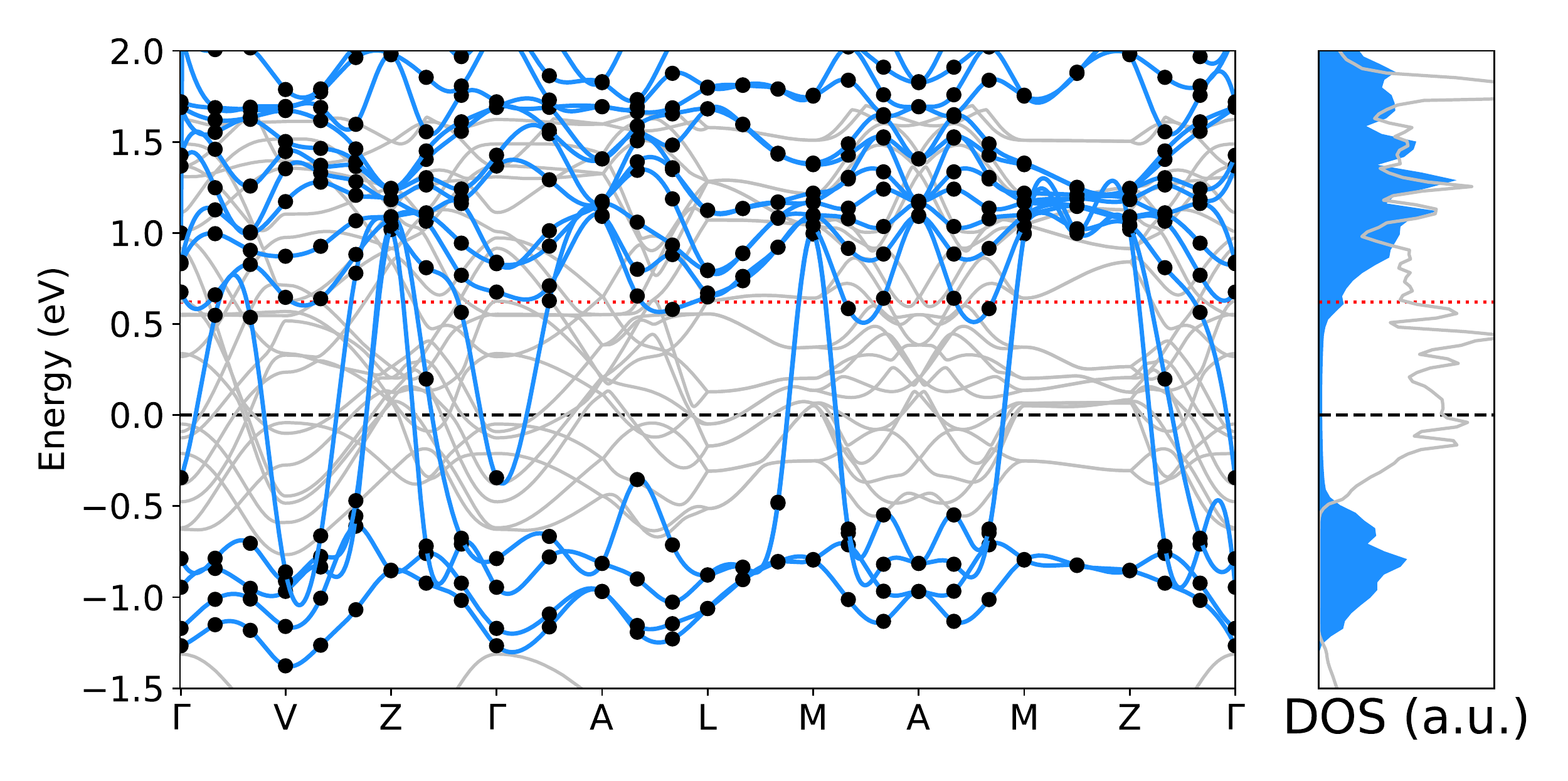}}
    \caption{\raggedright {DFT (blue lines) and PS-GW (filled circles) Bandstructures of a) M$_{2}$ VO$_2$ (this is a reproduction of the data originally published in Booth \textit{et al.}\cite{Booth2016a}) b) the ``M$_{2}$ Tetragonal" structure and c) the ``Peierls Paired" structure, in each case the Fermi level is set to 0 eV. and the PS-GW data is fitted with blue splines to guide the eye (not these are not true electron bands).}}
    \label{fig:DOSs}
\end{figure*}

\section{Methods}
\subsection{Structures}
The lattice parameters of the M$_2$ structure were obtained by Rietveld analysis of the X-ray Diffraction data of a sample of M$_{2}$ VO$_2$ prepared by deposition onto montmorillonite \cite{Booth2009} using the atomic coordinates of Marezio \textit{et al}.\cite{Marezio1971} The Tetragonal structure parameters used were those experimentally determined by Andersson.\cite{Andersson1954} The diffraction data were acquired at the Australian Synchrotron, using a beam energy of 15 keV, two detector offsets were used for the acquisitions and merged in post-processing. 

As Figure \ref{fig:schematics} indicates, the metal-insulator transition of M$_{2}$ VO$_2$ coincides with a displacive phase transition, which involves Peierls pairing of one half of the vanadium chains which run down the monoclinic b-axis, and an antiferroelectric distortion of the other (interleaved) chains. In order to examine the effects of these distortions on the electronic structure, intermediate structures were generated as follows. The ``Peierls Paired" structure consists of the M$_{2}$ structure with the antiferroelectric distortion removed, i.e. the AF chains in Figure \ref{fig:schematics}a are symmetrized such that they are evenly spaced and collinear as per those of the tetragonal structure. The ``M$_{2}$ Tetragonal" structure is generated by removing \textit{both} the antiferroelectric and Peierls distortions. This creates a structure in which the vanadium atoms are all evenly spaced and collinear, however their internuclear distances are slightly larger than the metallic tetragonal structure, and the structure retains the monoclinic $\beta$ angle of 91.88 $^\circ$. Thus, calculations of the electronic structure of the ``Peierls Paired" form illustrate the effect of introducing the Peierls pairing to the high temperature tetragonal structure, while the ``M$_{2}$ Tetragonal" structure explores the effect of expanding the inter-vanadium spacing, which occurs when the AF chain is created, but decouples the Peierls pairing and the increased bonding in the z-axes of the octahedra occuring \textit{via} the shifts in the (201) plane (see Figure 1c).

\subsection{Calculations}
The GW calculations were performed using the implementation of Shishkin and Kresse \cite{Shishkin2006,Shishkin2007} as contained in the Vienna Ab Initio Simulation Package (VASP),\cite{Kresse1996} after first calculating input wavefunctions using DFT\cite{Kohn1965} with GGA\cite{Perdew1996} functionals, on $4\times6\times6$ Monkhorst-Pack\cite{Monkhorst1976} \textbf{k}-space grids using the Brillouin zone integration approach of Bloechl \textit{et al}.\cite{Bloechl1994} Strong correlations were included by setting the derivatives of the wavefunctions with respect to the \textbf{k}-point grid to zero (the PS-GW method, see Booth \textit{et al.} \cite{Booth2016a}). This reduces the magnitudes of the overlap integrals of the polarizability matrix, $\chi(\textbf{q})$:\cite{Hybertsen1987}
\begin{multline}
\chi^{0}_{\mathbf{q}}(\mathbf{G,G^{\prime},\omega})=\frac{1}{\Omega}\sum\limits_{nn^{\prime}\mathbf{k}}^{}2w_{\mathbf{k}}(f_{n^{\prime}\mathbf{k+q}}-f_{n\mathbf{k}})\\\times
\frac{\langle\psi_{n^{'}\mathbf{k+q}}|e^{-i\mathbf{(q+G)r}}|\psi_{n\mathbf{k}}\rangle\langle\psi_{n\mathbf{k}}|e^{i\mathbf{(q+G^\prime)r^\prime}}|\psi_{n^{'}\mathbf{k+q}}\rangle}{\epsilon_{n^{\prime}{\mathbf{k}}+\mathbf{q}}-\epsilon_{n\mathbf{k}}-\omega+i\eta \text{sgn}[\epsilon_{n\mathbf{k}}-\epsilon_{n^{\prime}\mathbf{k}+\mathbf{q}}]}
\label{eq:chi}
\end{multline}
as well as the transitions back to the ground state as the bubble closes, as this is a modification of the Random Phase Approximation. Reducing the overlap integrals prevents the structure from polarizing to reduce electron correlations. Physically this is a manifestation that transitions out of the ground state will incur an energy penalty, and by extension the screening of an excited state must also scatter momentum states and will incur an energy penalty in the form of the Hubbard $U$ term as per the Hamiltonian:

\begin{equation}
H=-t\sum\limits_{\langle ij\rangle}^{}(c^{\dagger}_{i\sigma}c_{j\sigma}+c^{\dagger}_{j\sigma}c_{i\sigma})+\\U\sum\limits_{i}^{}n_{i\uparrow}n_{i\downarrow}
\end{equation}

This penalty can be simulated by reducing the overlap integrals to reduce the polarizability and thus including more of the bare Hartree-Fock interaction. The use of Projector Augmented Waves\cite{Blochl1994b,Shishkin2006} allows this increased interaction to be included as an on-site interaction as per the usual Hubbard Hamiltonian in the position basis above. For a more comprehensive presentation the reader is referred to Booth \textit{et al.}\cite{Booth2016a} Five self-consistency steps were used and all strongly correlated calculations were performed at a single frequency point, $\omega$=0.\cite{{Booth2016a}} In all calculations the Fermi level is set to zero energy.

\section{Results and Discussion}
\subsection{Peierls Distortion}
Figure \ref{fig:DOSs} presents the DFT (blue lines) and PS-GW (filled circles) band structures of the M$_{2}$ structure, the ``M$_{2}$ Tetragonal" structure and the ``Peierls Paired" structures respectively. As reported previously, strong correlations split the partially filled states of the M$_{2}$ structure into upper and lower Hubbard bands.\cite{Booth2016a} The ``M$_{2}$ Tetragonal" structure also exhibits considerable splitting of the states near $E_{F}$ into Hubbard bands, which is obviously difficult to confirm experimentally, given that this structure does not exist. This is to be expected however, given that the true tetragonal structure is itself strongly correlated,\cite{Tomczak2007} and thus does not exist at 0 K either. However, what it does provide is a way of determining which states move due to the two different structural distortions by comparing it to both the M$_{2}$ and ``Peierls Paired" structures.

\begin{figure}[h!]
    \begin{center}
    \subfigure[]{\includegraphics[width=0.7\columnwidth]{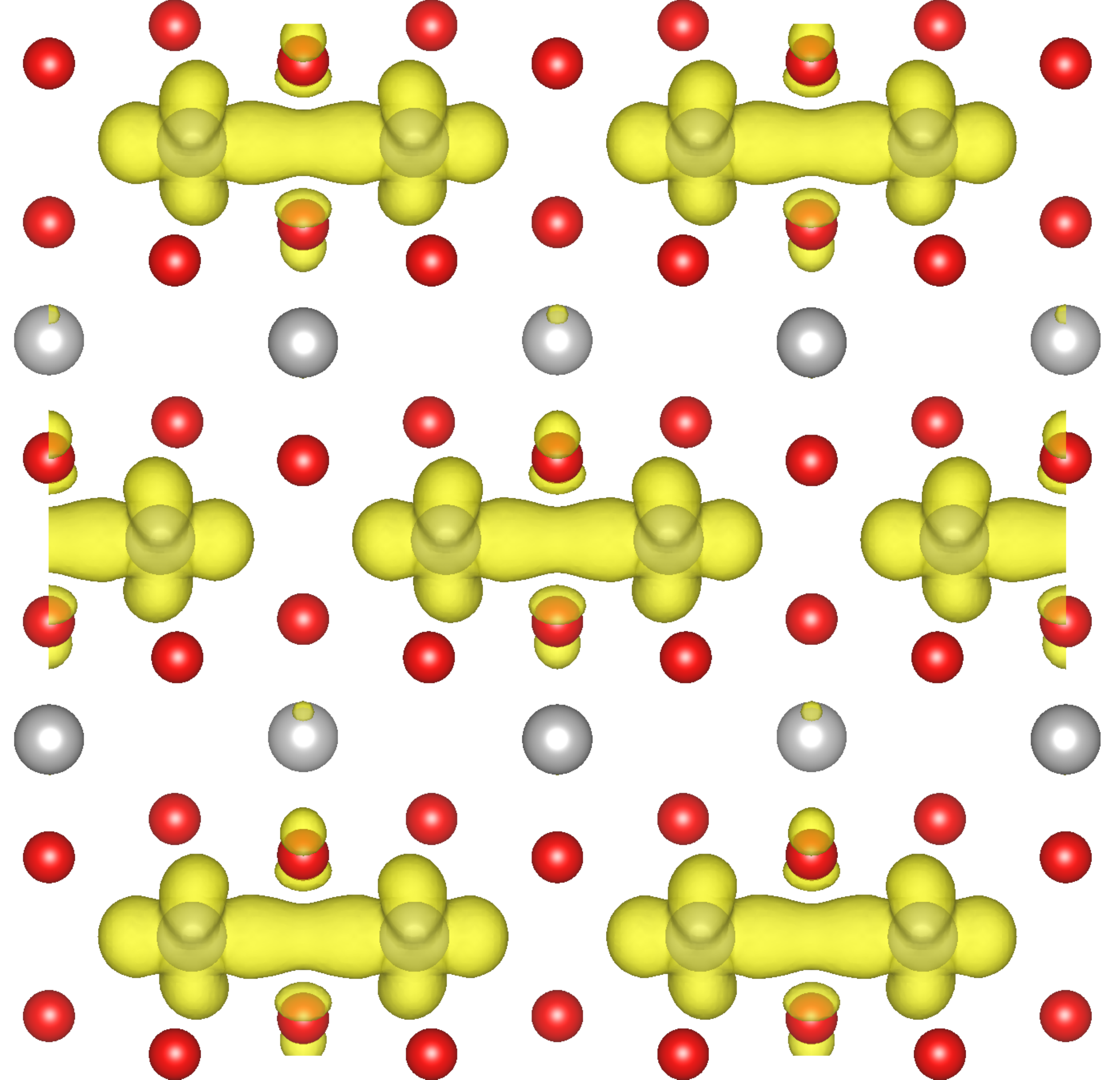}}\\
    \subfigure[]{\includegraphics[width=0.75\columnwidth]{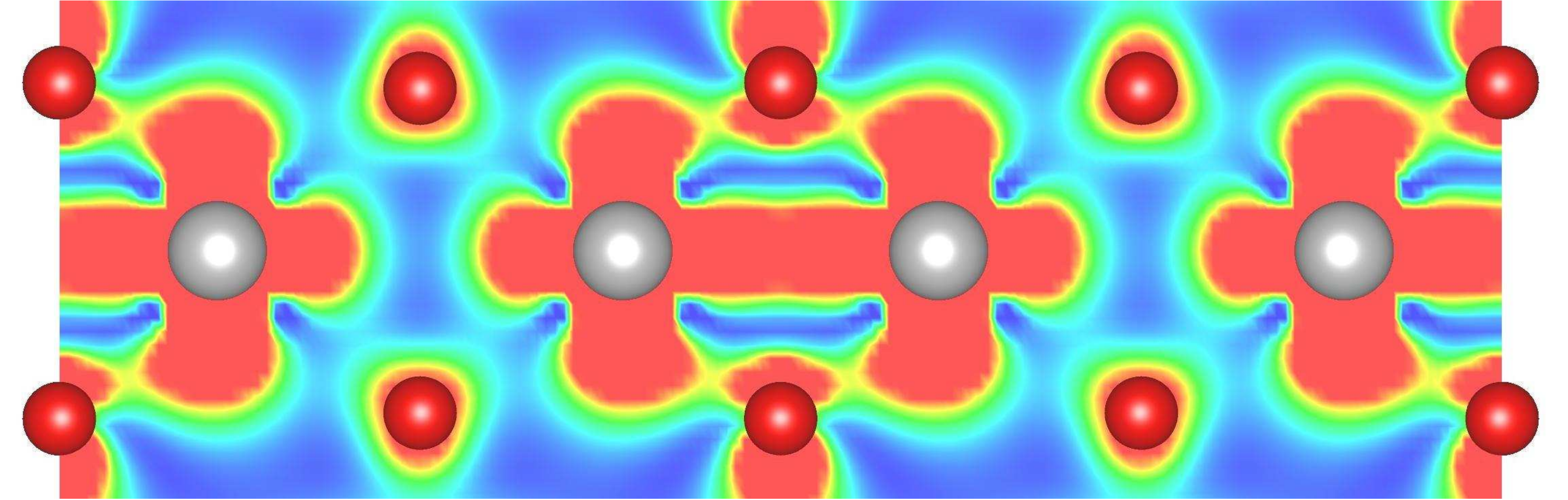}}\\
    \subfigure[]{\includegraphics[width=0.75\columnwidth]{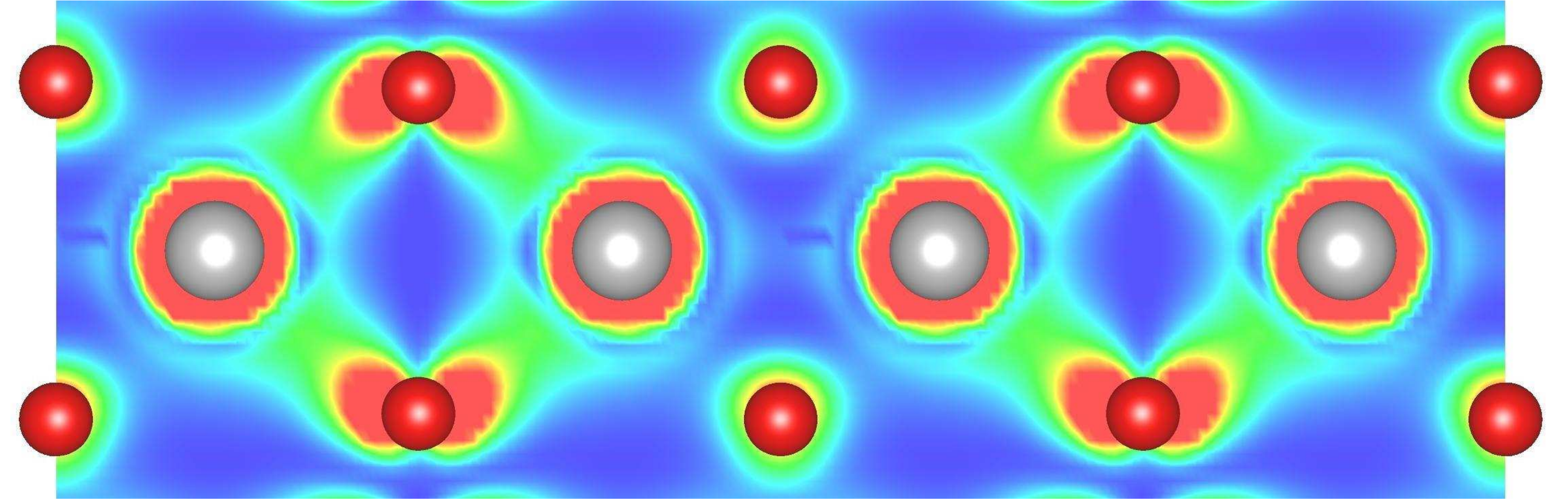}}
    \end{center}
    \caption{\raggedright{a) Charge density isosurface in the (201) plane of the $M \rightarrow Z$ region of the M$_{2}$ ``Peierls Paired" structure b) charge density slice of the M$_{2}$ structure in the (201) plane of the lower Hubbard band and c) charge density slice of the upper Hubbard band, suggesting poor overlap due to the different interstitial charge densities.}}
    \label{fig:CD1}
\end{figure}
A comparison of Figures \ref{fig:DOSs}a and \ref{fig:DOSs}c with figure \ref{fig:DOSs}b reveals that the Peierls distortion results in considerable stabilisation of states in the $M \rightarrow Z$ direction in the lower Hubbard band. In the ``M$_{2}$ Tetragonal" structure, the state at $M$ sits in the upper Hubbard band, while the subsequent states are below $E_{F}$, indicating that the associated band crosses from the upper to the lower Hubbard band. However, in both the M$_{2}$ and the ``Peierls Paired" structures, the $M \rightarrow Z$ region consists of a flat band sitting well below $E_{F}$ in the lower Hubbard band. Therefore, the imposition of the Peierls pairing has stabilized these states, dropping the band below $E_{F}$ across this region of k-space. We can get an idea of why this occurs by plotting a charge density isosurface corresponding to these points (Figure \ref{fig:CD1}a).

The isosurface of this charge density in the (201) plane indicates that these ``stabilized" states correspond almost entirely to charge density concentrated on the Peierls paired vanadium atoms. The charge density also extends across the interstitial region between the short V-V distance, creating bonding density between the Peierls pairs. Thus, from Figures \ref{fig:DOSs} and \ref{fig:CD1}a the effect of imposing Peierls pairings on the tetragonal structure is the formation of bonding density, pulling the bonding states below $E_{F}$, with a corresponding destabilisation of the antibonding states. 
\begin{figure}[h!]
    \begin{center}
    \subfigure[]{\includegraphics[width=0.6\columnwidth]{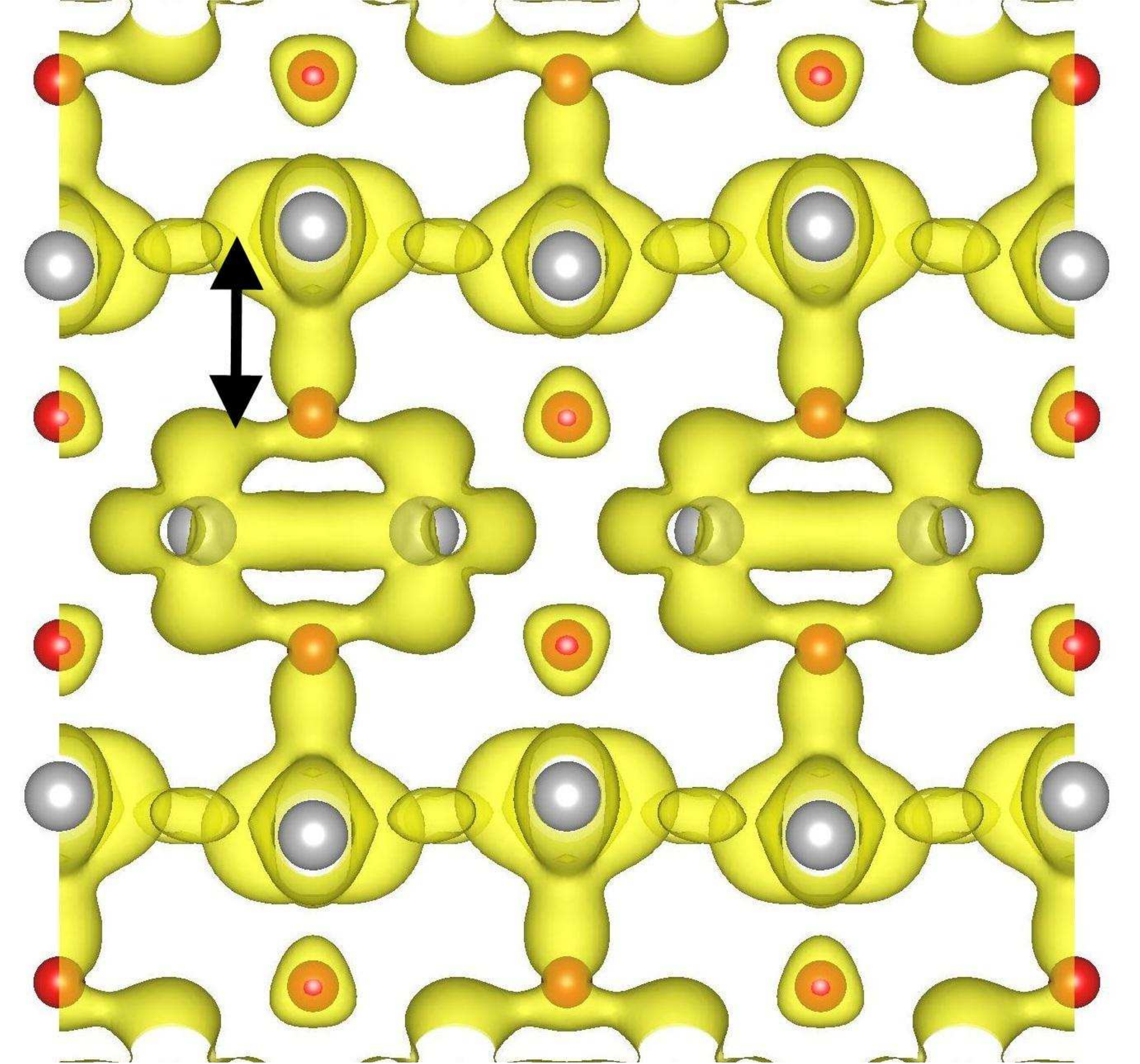}}
    \subfigure[]{\includegraphics[width=0.6\columnwidth]{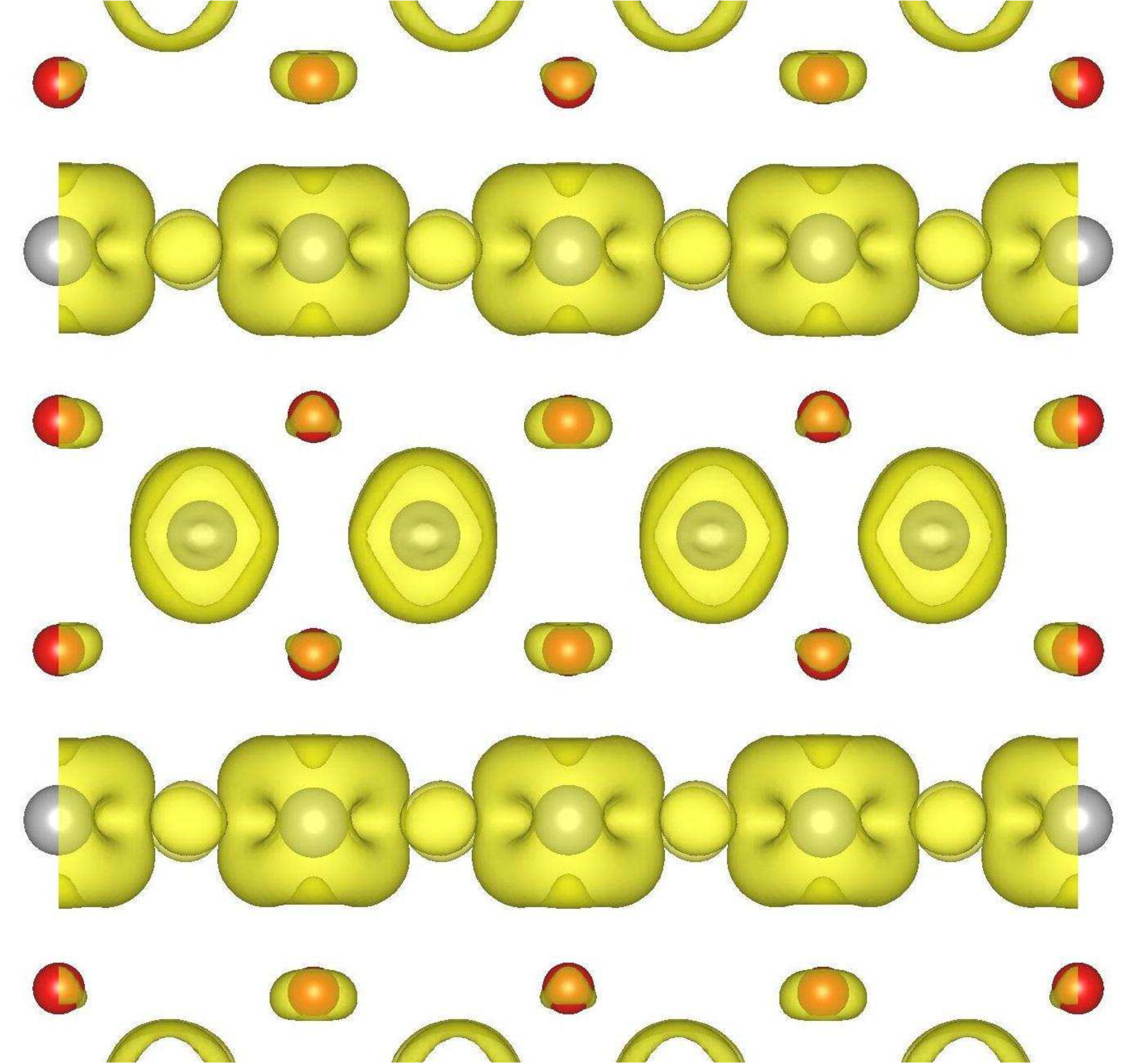}}\\
    \end{center}
    \caption{\raggedright {a) Isosurface of the charge density in the (201) planes of the lower Hubbard band of the M$_{2}$ structure (in which the increased V-O distance from the AF distortion is marked with an arrow) and b) corresponding isosurface of the the ``Peierls Paired" structure.}}
    \label{fig:CD2}
\end{figure}

\begin{figure*}[h!]
    \centering
    \subfigure[]{\includegraphics[width=0.8\columnwidth]{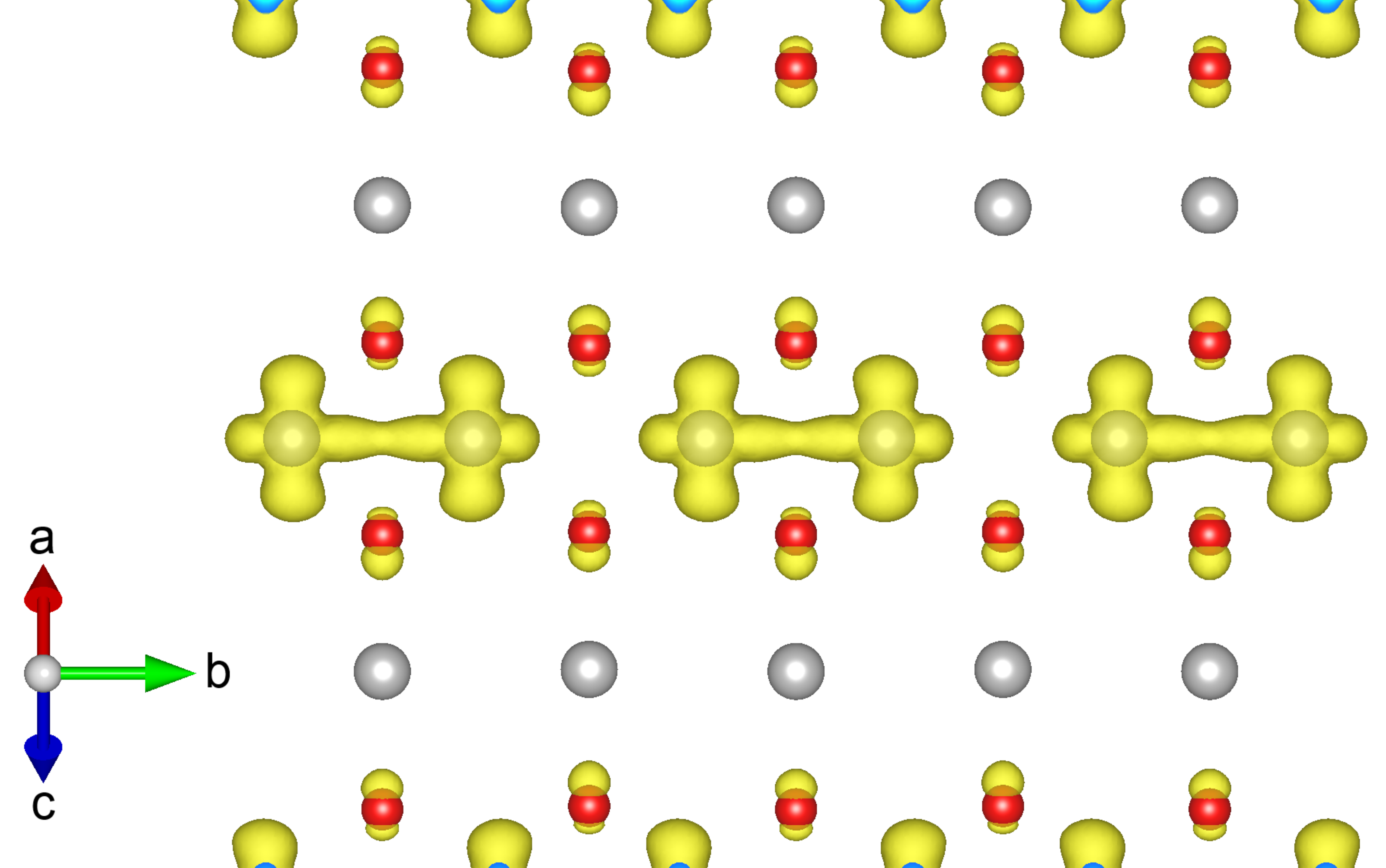}}
    \subfigure[]{\includegraphics[width=0.8\columnwidth]{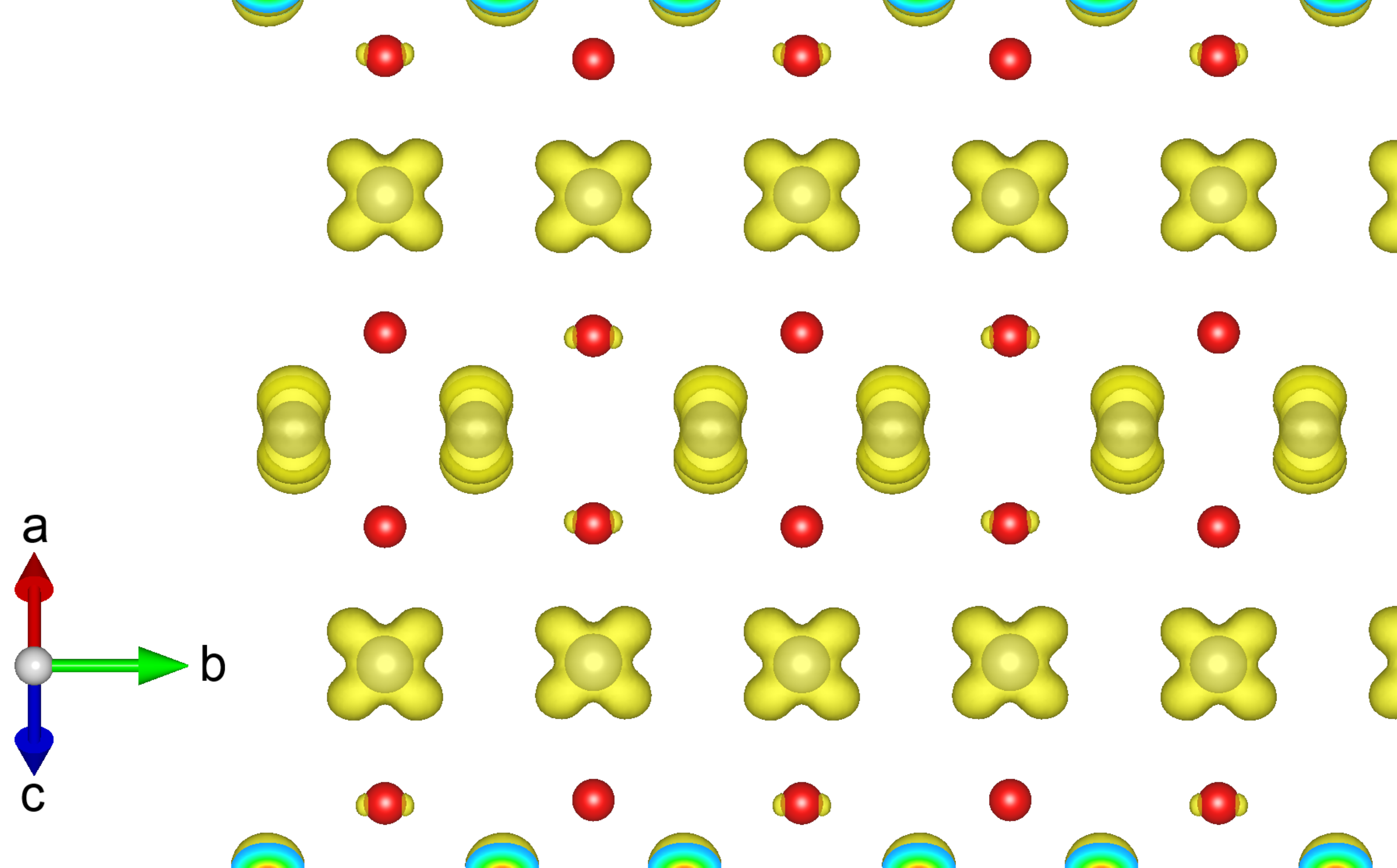}}\\
    \subfigure[]{\includegraphics[width=0.8\columnwidth]{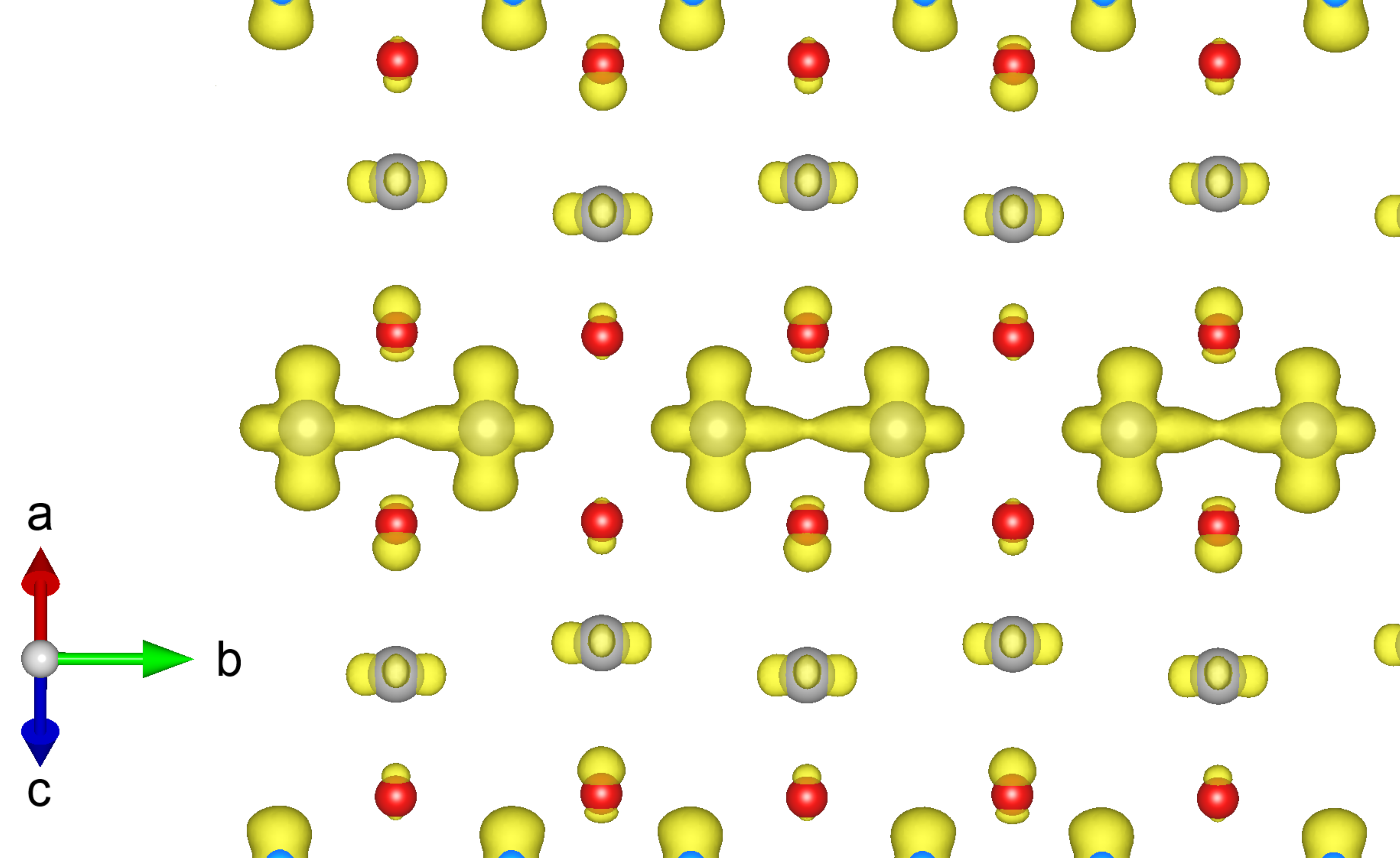}}
    \subfigure[]{\includegraphics[width=0.8\columnwidth]{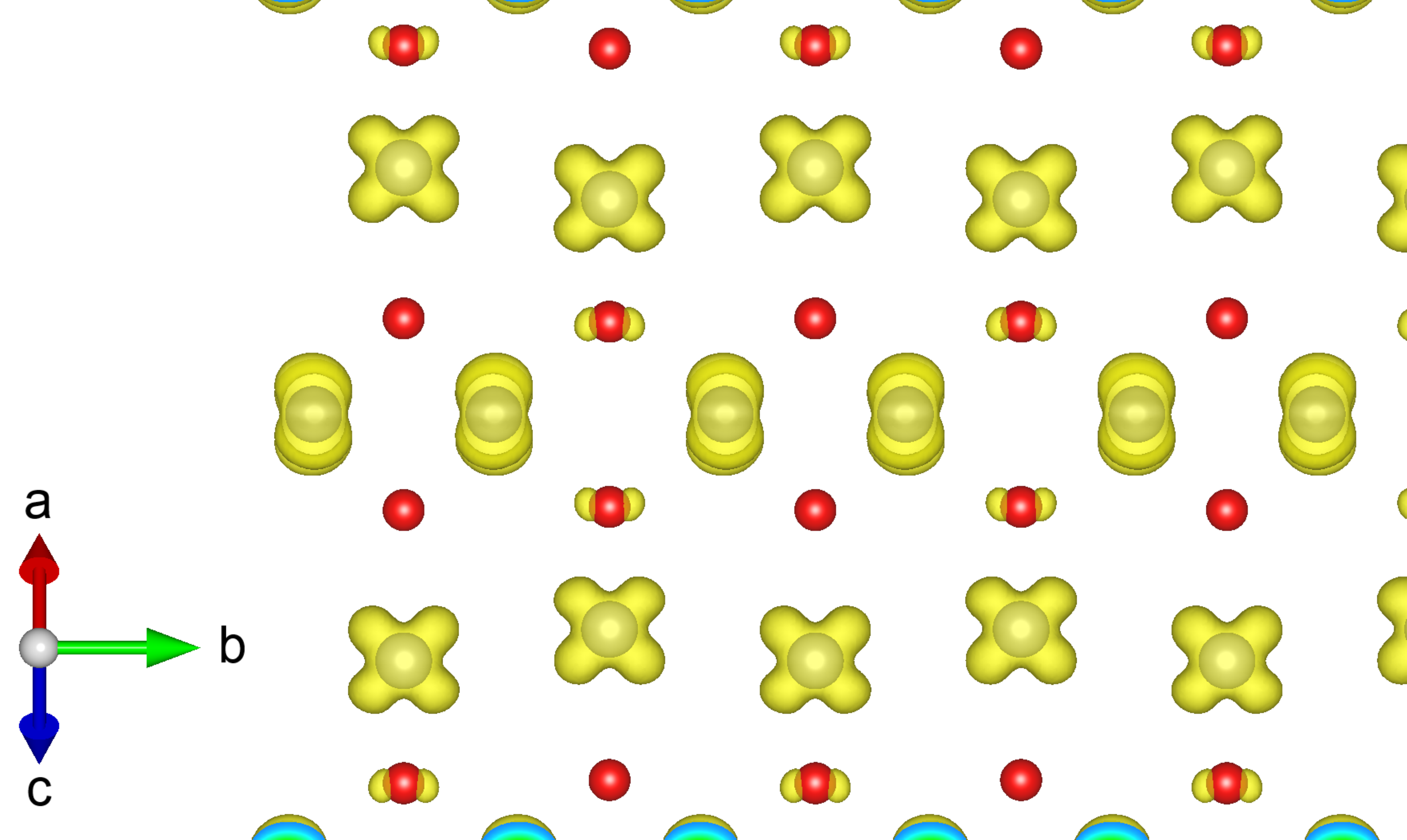}}\\
    \subfigure[]{\includegraphics[width=1.2\columnwidth]{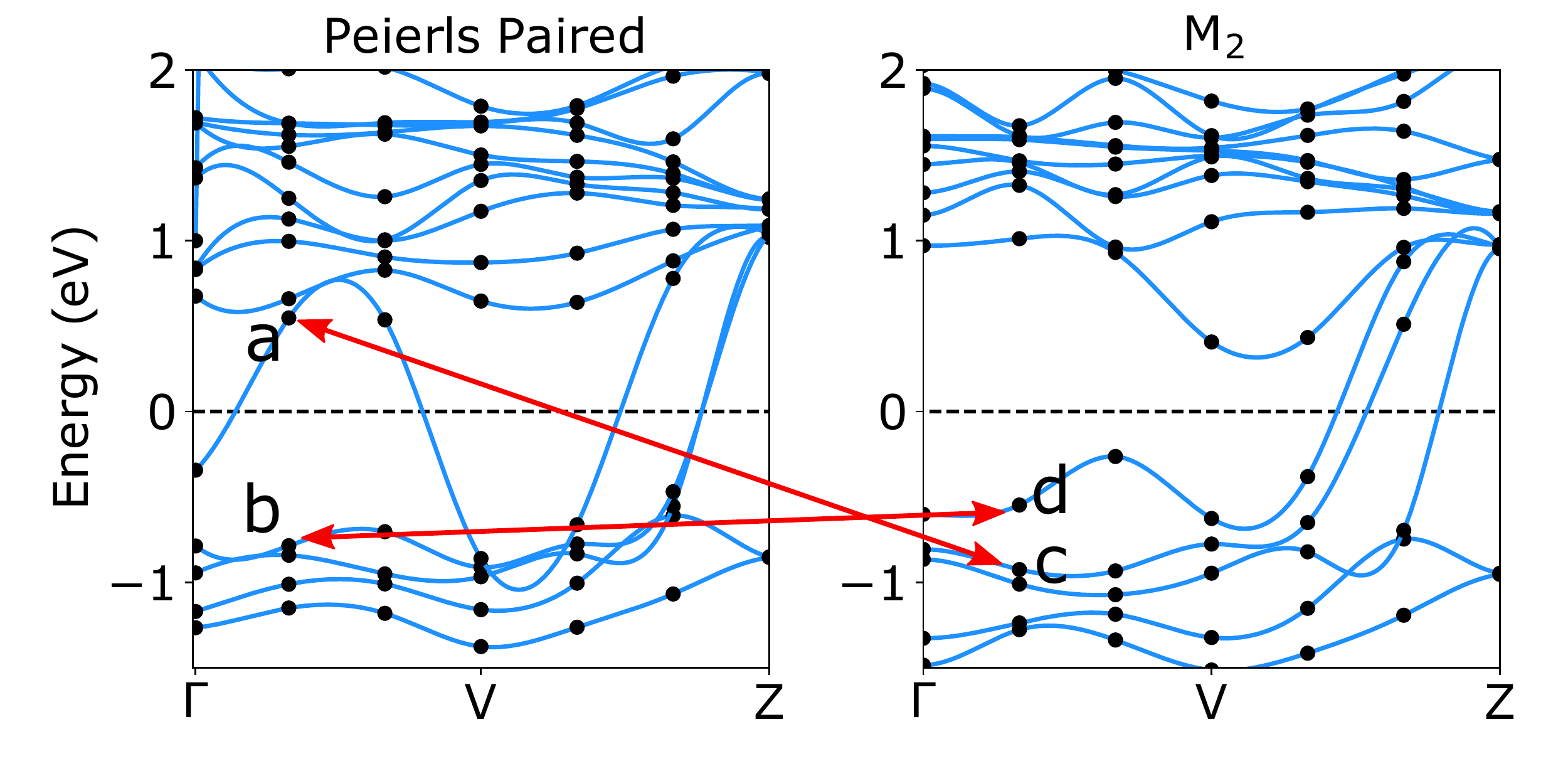}}
    \caption{\raggedright{a) Charge density in the (201) plane of the two states marked with red circles in Figure \ref{fig:BZ}} in the upper Hubbard band of the ``Peierls Paired" structure, b) corresponding charge density of the same states but in the lower Hubbard band, c) charge densities of the same wavefunctions as per a) but this time of the M$_{2}$ structure, d) charge densities of the same wavefunctions as b) but this time in the M$_{2}$ structure, e) band structures of the ``Peierls Paired" and M$_{2}$ structures in the $\Gamma \rightarrow V \rightarrow Z$ region with the states corresponding to the charge densities of a)-d) indicated by the corresponding letters, which indicates that the imposition of the antiferroelectric distortion causes the bonding states of a) to cross the gap and sit in the lower Hubbard band.}
    \label{fig:crossover}
\end{figure*}

In addition to the localization induced by the increased nuclear potential overlap, the breaking of the tetragonal symmetry will result in a decrease in the exchange charge density between points in k-space corresponding to bonding and anti-bonding states. Figures \ref{fig:CD1}b-c illustrate this using charge density slices of the lower and upper Hubbard bands in the (201) plane respectively. As the atoms are paired, the previously symmetric states split into bonding (Figure \ref{fig:CD1}b) and anti-bonding (Figure \ref{fig:CD1}c) combinations. The overlap of these wavefunctions will obviously decrease given the different forms of the charge density on the Peierls paired chain. Therefore, the overlap integrals corresponding to transition between these states in the polarizability matrix (equation \ref{eq:chi}) will be smaller. This results in more of the bare Hartree-Fock interaction being included, increasing correlations and splitting states at the Fermi level.

Figure \ref{fig:CD2} illustrates the charge density of the lower Hubbard bands of the M$_2$ structure and the ``Peierls Paired" structure respectively, and it is apparent that the lower Hubbard band of the M$_2$ structure is comprised of a significant number of states which contain bonding density between the Peierls Paired vanadium atoms, while the ``Peierls Paired" structure's lower Hubbard band contains non-bonding density on the Peierls chain. Therefore, while imposing the Peierls Pairing may stabilize the states of the $M \rightarrow Z$ region, it does not result in a significant number of such states inhabiting the lower Hubbard band in the ``Peierls Paired" structure.

Searching for an origin of this discrepancy, we see that by comparing the band structures of Figures \ref{fig:DOSs}b-c, in addition to the stabilization of the states in the $M \rightarrow Z$ region, a peak develops in the density of states of the ``Peierls Paired" structure, corresponding to states in the $\Gamma \rightarrow V$ and $M \rightarrow A$ regions. Figure \ref{fig:DOSs}c indicates this with a horizontal red line at the peak in energy in the band structure and DOS plots. Figure \ref{fig:crossover} explores the nature of these states in detail. Figure \ref{fig:crossover}a is a charge density isosurface of the states indicated with red circles in Figure \ref{fig:BZ} which form part of the peak just below the upper Hubbard band in the ``Peierls Paired" structure. Figure \ref{fig:crossover}b plots a charge density isosurface of the same points in \textbf{k}-space but these states are from the band below that of Figure \ref{fig:crossover}a, and inhabit the lower Hubbard band. Comparing these two we see that the bonding density, of the type seen in the lower Hubbard band in the $M \rightarrow Z$ region of both the M$_{2}$ and ``Peierls Paired" structures in these regions sits just below the upper Hubbard band and is thus much higher in energy, while the non bonding density sits in the lower Hubbard band. Figures \ref{fig:crossover}c-d present isosurfaces of the charge density of the same wavefunctions for the M$_{2}$ structure, and comparing these with Figure \ref{fig:crossover}a-b it is obvious that they are virtually identical with the slight change in structure resulting in a shift of some of the bonding density on the Peierls paired chain to the antiferrelectrically distorted chain. However the energies are very different. 

Figure \ref{fig:crossover}e plots the band structures of the ``Peierls Paired" and M$_{2}$ structures in the $\Gamma \rightarrow V \rightarrow Z$ region, and the states corresponding to the charge densities of Figures \ref{fig:crossover}a-d are indicated with their corresponding letters. From this plot we see that while the ``Peierls Paired" structure has dropped the energies of the states with bonding density to just below the upper Hubbard band, adding the antiferroelectric distortion of the other vanadium chain, which takes it into the M$_{2}$ structure, drops these states into the lower Hubbard band, and in fact they cross over the non-bonding band to sit at lower energy. Therefore, the effect of the antiferroelectric distortion is to considerably stabilize the bonding states. Note that each of the charge densities are a sum over the two \textbf{k}-points highlighted in Figure \ref{fig:BZ}, and both states exhibit this crossover, but for presentation reasons only one is focused on in Figure \ref{fig:crossover}e. 

\begin{figure*}[h!]
\centering
\subfigure[]{\includegraphics[width=0.6\columnwidth]{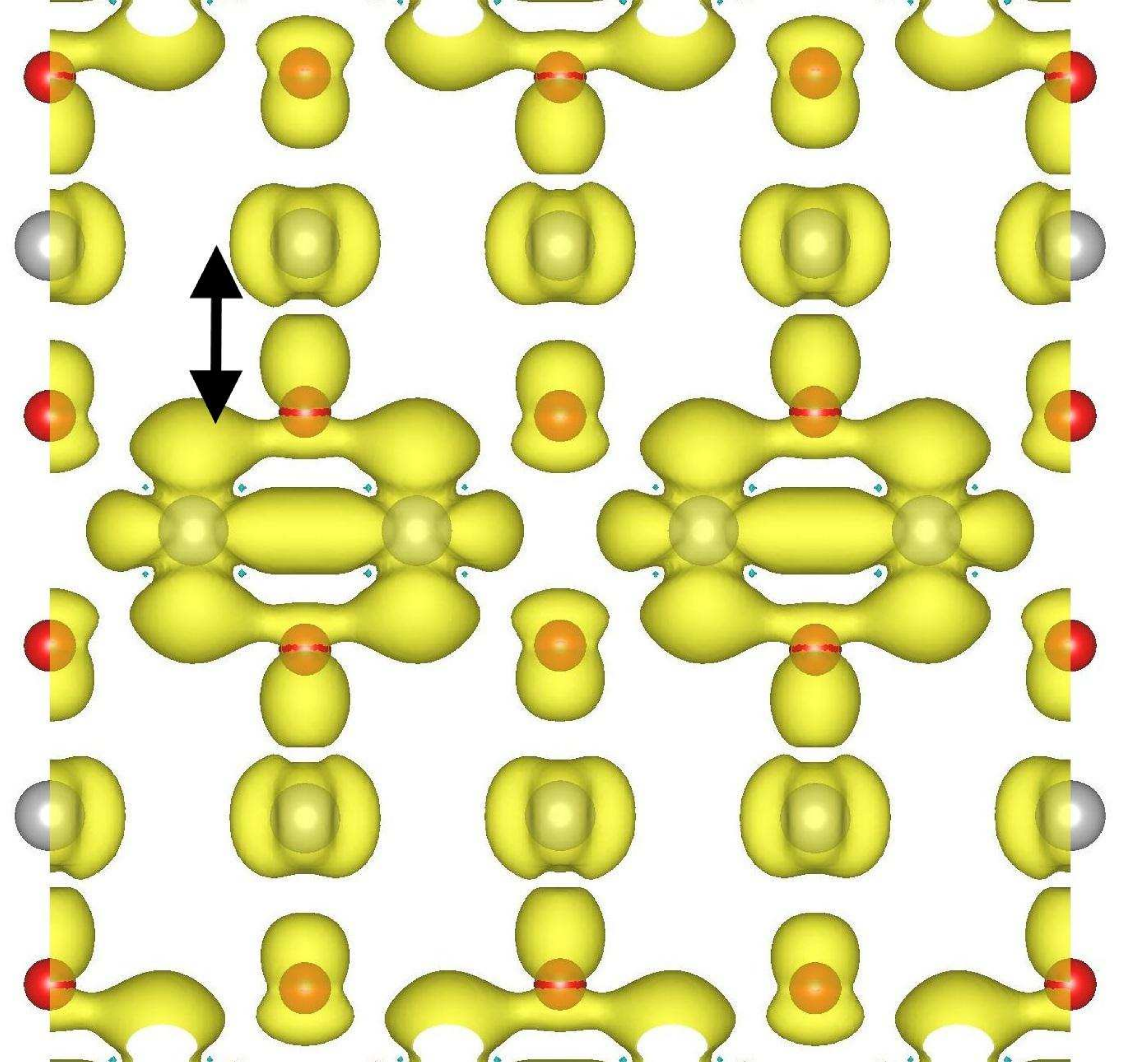}}
\subfigure[]{\includegraphics[width=0.8\columnwidth]{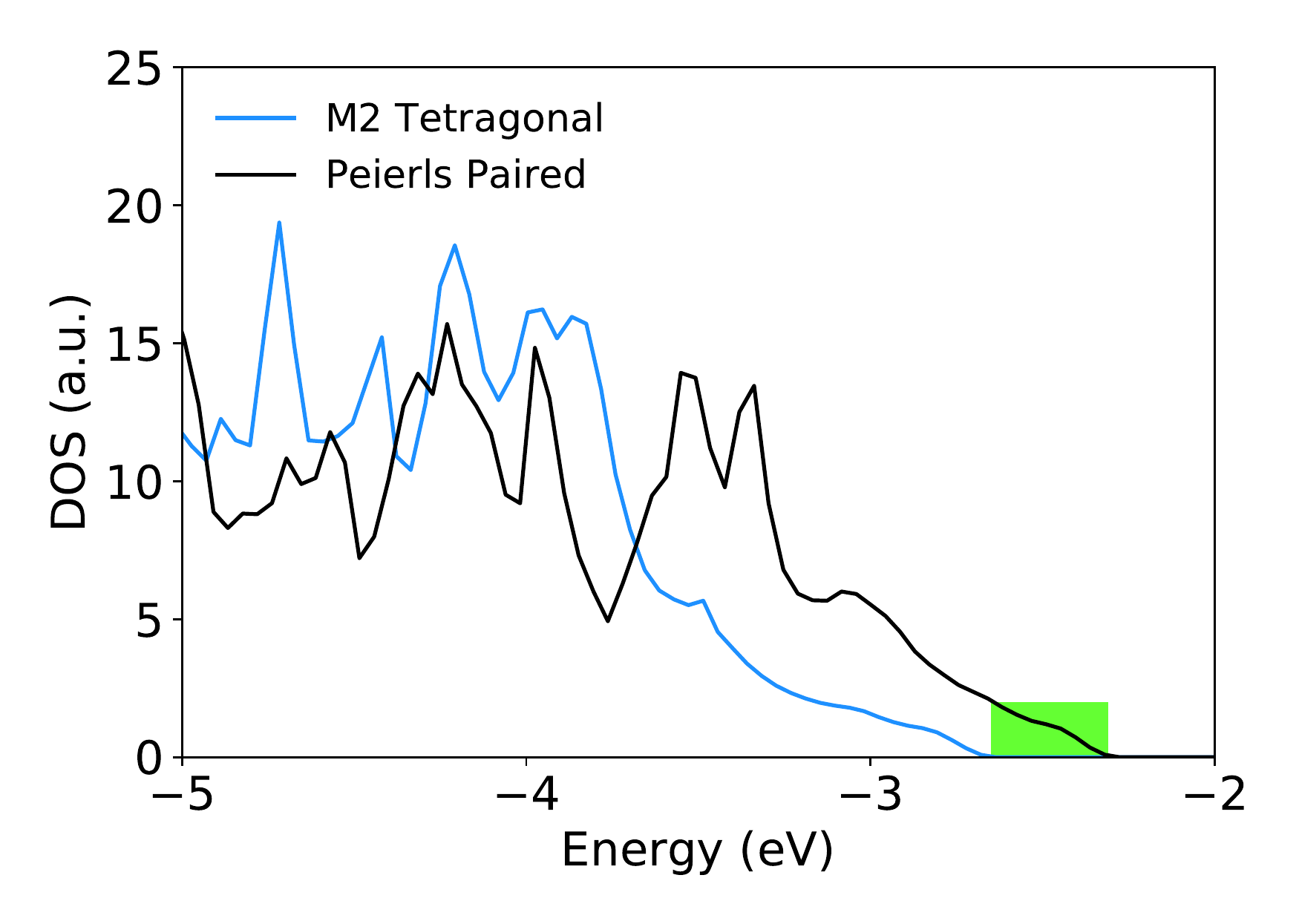}}\\
\subfigure[]{\includegraphics[width=0.6\columnwidth]{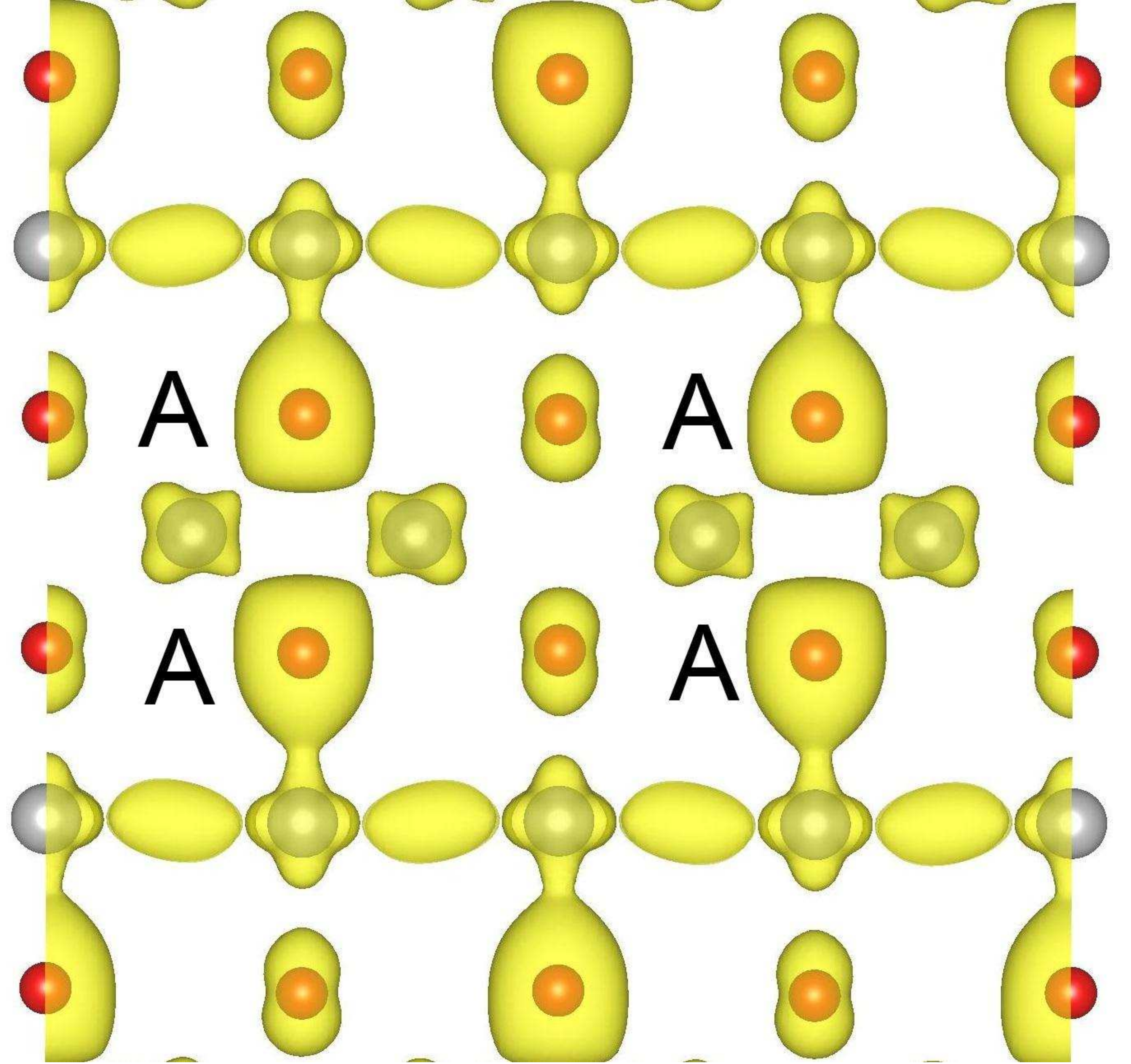}}
\subfigure[]{\includegraphics[width=0.8\columnwidth]{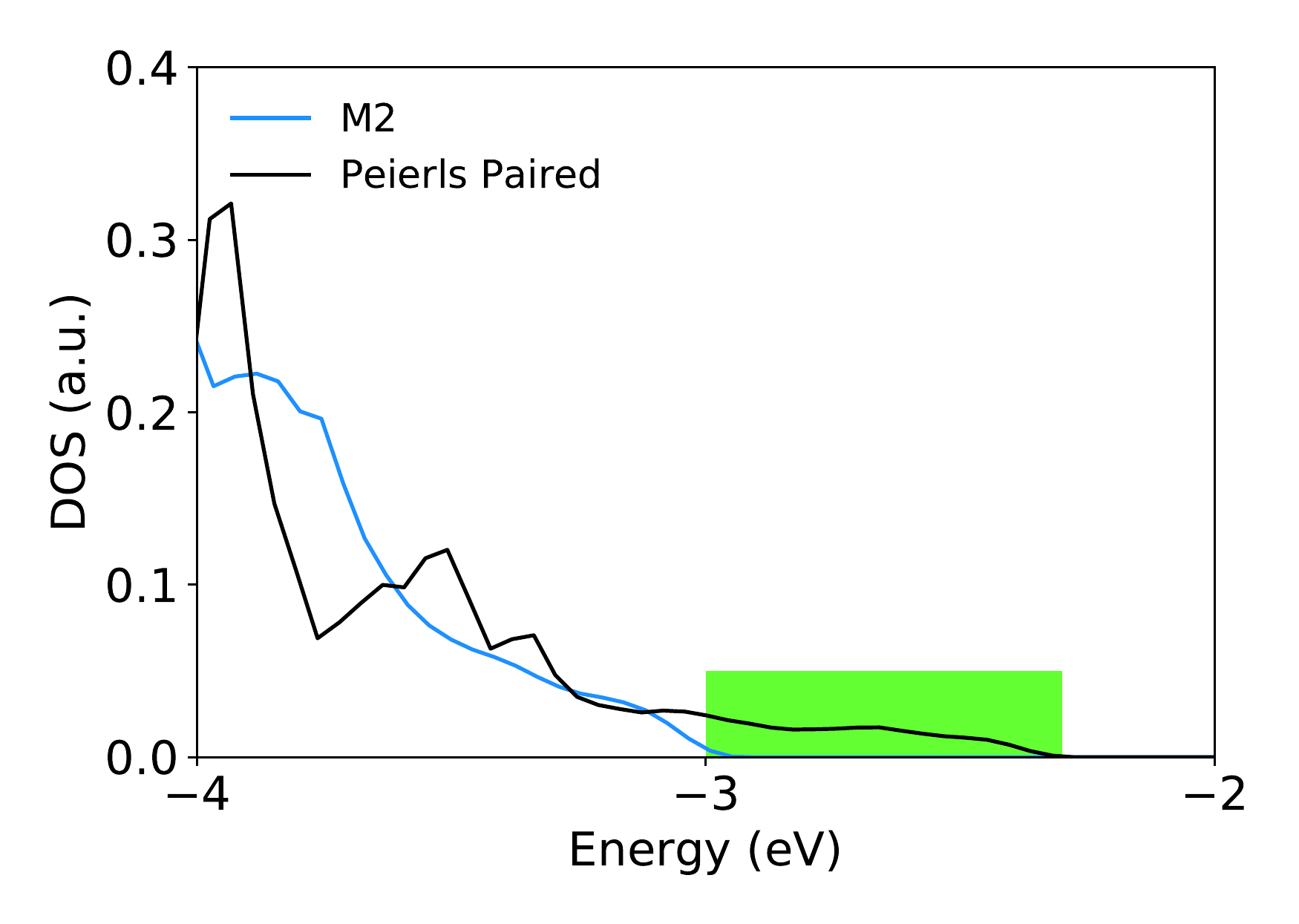}}
\caption{\raggedright {a) charge density in (201) states corresponding to the peak in the DOS of Figure \ref{fig:DOSs}c marked by the red line, the symmetrized V-O distance is marked by an arrow, b) comparison of the densities of states of the oxygen bands of the ``M$_{2}$ Tetragonal" and ``Peierls Paired" structures, c)charge density of the total DOS corresponding to the shaded energy region of b), where the apical oxygen atoms are marked ``A", d) partial densities of states of the apical oxygen atoms of the M$_{2}$ and Peierls paired structures.}}
\label{fig:Oxygen}
\end{figure*}

\subsection{Antiferroelectric Distortion}
While Figures \ref{fig:DOSs} and \ref{fig:crossover} illustrate that in the ``Peierls Paired" structure bonding is not enough to drop the states into the lower Hubbard band, it is easy to reconcile this by examining the changes in bond lengths introduced by the antiferroelectric distortion. In the M$_{2}$ structure the internuclear distance between the apical oxygen and the antiferroelectrically distorted vanadium atoms is 2.12 \AA, while in the ``Peierls Paired" structure it is 1.915 {\AA}. Figure \ref{fig:Oxygen}a plots a charge density isosurface of the ``Peierls paired" structure in the (201) plane, and marks the distance from the vanadium atoms to the apical oxygen atom of the AF chain with a double headed arrow. Comparing this to the same distance of the M$_{2}$ structure in Figure \ref{fig:CD2}a (also marked with a double-headed arrow) we see that in addition to it being shorter, the charge density is now not connected to the oxygen atom, despite the isoasurface levels being identical. 

Inspecting the densities of states of corresponding to the oxygen bands of the ``M$_{2}$ Tetragonal" and the ``Peierls Pared" structures in Figure \ref{fig:Oxygen}b, it can be seen that imposing the Peierls pairing shifts the leading edge of the density of states upward in energy, thus they become less stable (green highlighted region). Figure \ref{fig:Oxygen}c transforms this highlighted region to charge density, and reveals that as expected, this density is mostly concentrated on the apical bridging oxygen atoms, marked by the letter A. Figure \ref{fig:Oxygen}d plots the partial densities of states of the apical oxygen atoms of the ``Peierls Paired" structure and the M$_2$ structure, and the ``Peierls Paired" structure clearly exhibits the same shift observed in the total density of states of Figure \ref{fig:Oxygen}b (again highlighted in green). This shift is only observed on the apical oxygen atoms however, confirming that it is due to the decrease in the V-O distance. Comparing this with the M$_{2}$ partial density of states for the apical oxygen atoms reveals that imposing the antiferroelectric distortion stabilizes these states. Therefore increasing the charge density in this region by imposing the Peierls pairing, without a corresponding antiferroelectric distortion on the other vanadium chain will result in $d$-electrons experiencing stronger repulsion which, as Figure \ref{fig:DOSs}c  indicates, almost completely counteracts the decrease in energy from the Peierls pairing creating bonding configurations. Increasing the V-O bond distance by introducing the anti-ferroelectric distortion lowers the energies of the oxygen states, dropping them back into the broad oxygen $p$-band. Thus the antiferroelectric distortion of the AF chain is revealed as simply a consequence of electrostatic repulsion generated by the Peierls pairing, which also increases the V-V distance along the chain, expanding the unit cell.

This also reveals the unusual stabilization of the bonding states in the $M \rightarrow Z$ region of the Peierls Paired structure (Figure \ref{fig:CD1}a); these states contain almost no charge density on the apical oxygen atoms. Therefore this electrostatic repulsion is minimal, and the states sit in the lower Hubbard band.

\section{Conclusion}
Putting all of this together, the data confirms that the Peierls pairing observed, like that of the M$_{1}$ structural phase transition, produces bonding/antibonding splitting and suggests that the dimerization is an attempt at localization. The concurrent antiferroelectric distortion in the other half of the vanadium chains is merely due to the minimization of the repulsion between the electrons on the vanadium atoms and negatively charged apical oxygen atoms. In the M$_{1}$ structure the same pairing and antiferroelectric distortion manifests, however \textit{both} chains pair. Given that the M$_{2}$ structure is commonly observed in doped vanadium dioxide systems,\cite{Marezio1971,Booth2009} a possible reason for the transition from M$_{1}$ to M$_{2}$ upon doping is that the doped ions, which contain either fewer (such as Ti\cite{Booth2009}) or more electrons (such as Cr\cite{Marezio1971}) disrupt the Peierls pairing and thus inhabit the antiferroelectrically distorted chains. This configuration allows the structure to isolate the sites with unpaired electrons in the usual Mott manner by increasing the inter-atomic spacing, while the other chain consisting of vanadium atoms Peierls pairs. This work thus confirms that M$_{2}$ VO$_{2}$ is Mott insulating, but also reveals how the electronic states are rearranged by the structural motions.

\section{Acknowledgements}
JMB and SPR acknowledge the support of the ARC Centre of Excellence in Exciton Science (CE170100026). DWD acknowledges the support of the ARC Centre of Excellence for Nanoscale BioPhotonics (CE140100003). This work was supported by computational resources provided by the Australian Government through the National Computational Infrastructure and the Pawsey Supercomputer Centre. 
\bibliography{C:/Local_Disk/GWApproximation/Bibliography/library}
\end{document}